\documentclass[prb,twocolumn,showpacs,showkeys,amssymb]{revtex4}

\usepackage{graphicx}
\usepackage{dcolumn}
\usepackage{bm}

\begin{document}

\title{Mean-field dynamics of a quantum dot - microcavity system}

\author{Herbert Vinck-Posada$^{(1)}$, Boris A. Rodriguez$^{(1)}$,
 Augusto Gonzalez$^{(1,2)}$}

\affiliation{$^{(1)}$Instituto de Fisica, Universidad de Antioquia, AA 1226,
 Medellin, Colombia\\
 $^{(2)}$ Instituto de Cibernetica, Matematica
 y Fisica, Calle E 309, Vedado, Ciudad Habana, Cuba}

\begin{abstract}
Mean-field evolution equations for the exciton and photon populations and
polarizations (Bloch-Lamb equations) are written and numerically solved in
order to describe the dynamics of electronic states in a quantum dot coupled
to the photon field of a microcavity. The equations account for
phase space filling effects and Coulomb interactions among carriers,
and include also (in a phenomenological way) incoherent pumping of
the quantum dot, photon losses through the microcavity mirrors, and
electron-hole population decay due to spontaneous emission of the dot.
When the dot may support more than one electron-hole pair, asymptotic
oscillatory states, with periods between 0.5 and 1.5 ps, are found almost
for any values of the system parameters.
\end{abstract}


\maketitle

\section{Introduction}

The coupling of light with an atomic state depends basically on two factors:
(a) the spread of the wave function or dipole moment of the atomic state, and
(b) the density of photon states in the energy region of interest. In the search
for optimizing this coupling, atomic Rydberg states and three-dimensional
microcavities, where the photon modes exhibit a discrete spectrum, have been
used. One of the salient realizations is the construction of micromasers
\cite{micromaser}.

More recently, excitonic states in quantum dots have proven to be excellent
candidates to replace atomic Rydberg states in a microcavity.
From one side, quantum dots can easily be incorporated inside the microcavity
by means of modern semiconductor technology, thus avoiding the complicated
step in which a beam of atoms in Rydberg states is prepared. On the other hand,
the effective Bohr radius in a semiconductor is a factor of $m_0\epsilon/(m\epsilon_0)$
larger than the corresponding radius in vacuum, where $m_0$ and
$\epsilon_0$ are, respectively, the electron mass and dielectric constant of
vacuum, and $m$, $\epsilon$ are the corresponding magnitudes in the semiconductor.
In GaAs, for example, this factor is around 200. As a result, the ground-state
exciton wave function is as spread as an atomic Rydberg state. Transitions
energies are typically in the infrared range.

In the present paper, we study the dynamics of a quantum dot-microcavity
system, focusing on the stationary or asymptotic photonic states in the
cavity. A previous, very serious, study of this dynamics \cite{Tejedor} has been motivated
by recent experimental work \cite{X1,X2,X3}, and theoretical papers \cite{XX1,XX2},
in which the exciton-cavity coupling is characterized, and the system is shown
to serve as an efficient source of single triggered photons, and a possible
source of entangled photons.

The results of Ref. \onlinecite{Tejedor} rely on the assumption that the quantum
dot may support only one excitonic state (a shallow dot). We relax this assumption,
allowing more than one excitonic state in the dot. Coulomb interactions among particles
are shown to play an important role in the dynamics, shifting the single-pair levels,
modifying the Rabi frequencies, and causing interference effects. The most relevant
consequence of including additional pair states in the dot, however, is shown to be
the existence of asymptotic oscillatory states, in which the number of photons
oscillates around a mean value, with periods between 0.5 and 1.5 ps (for the set of parameters used in the calculations).
We keep the assumption that the excitonic states are coupled to a single photon mode.
This means that the separation between photon modes in the cavity should be larger
than the splitting between excitonic states in the dot. As the latter magnitude is
of the order of a few meV, our results will be valid for cavities with a radius of
0.5 $\mu$m or smaller \cite{radius}.

A second, simplifying, working hypothesis is the mean field approximation. For electrons,
it means the neglect of correlation terms in the evolution (Bloch) equations. For
light, we use a semiclassical or coherent (Lamb) description, in which the field is
characterized by an amplitude and a phase. Let us stress that, in the intervals of
parameters we consider, when the number of photons proves to be significantly
different from zero the statistics of the photon field is Poissonian or near
Poissonian, at least in the one-exciton dot \cite{Tejedor}, in such a way that the
coherent approximation for the photon field is justified. The reward from this
simplification is that the system of evolution equations is relatively small,
allowing a qualitative analysis of its solutions, and the extension to deeper dots,
which may support more than one electron-hole states.

The plan of the paper is as follows. In Sec. \ref{sec2}, the basic evolution (Bloch-Lamb)
equations are obtained, and the initial conditions used to solve them are discussed.
Sec. \ref{sec3} is devoted to the presentation of results. The one-, two-, and
three-states dots coupled to a single photon mode (with a given polarization) are
studied. Additionally, the two-states dot coupled to a polarization degenerated light
mode is also considered. Finally, concluding remarks are given in Sec. \ref{sec4}.

\section{The theory}
\label{sec2}

Our model microcavity is assumed to have an ideal cylindrical shape. Its radius is
lower than 0.5 $\mu$m, in such a way that the spacing between photon modes is of
the order of 20 meV.\cite{radius} We will be interested in one of these modes. To
fix ideas, we speak about the lowest mode, polarization degenerated, which is
called HE$_{11}$ according to the terminology for modes in a guide \cite{modos}.
In the paper, we will conventionally denote the two polarization components as
$\sigma_{(+)}$ and $\sigma_{(-)}$.

A quantum dot is located at a point along the cavity axis in which the electric
field has a maximum. We assume a GaAs dot. The dot has also a cylindrical shape. Its height
is lower than 10 nm, in such a way that light-hole sub-bands have higher energy than the heavy-hole ones and, in a first approximation, could be ignored in the dynamics. A second reason for not including light holes is that their coupling to the photon modes is a factor of 3 less intense than the heavy hole coupling \cite{Bastard}.

Electron-hole pairing is induced by both Coulomb interactions and the coupling to the
photon mode. Thus, in addition to the electron, hole and photon populations:

\begin{eqnarray}
\rho^{(e)}_{nn} &=& \langle e_n^\dagger e_n\rangle,
 \label{eq1}\\
\rho^{(h)}_{\bar n\bar n} &=& \langle h_{\bar n}^\dagger h_{\bar n}\rangle, \\
|\sigma_{(+)}|^2 &=& \langle a_{(+)}^\dagger a_{(+)}\rangle, \\
|\sigma_{(-)}|^2 &=& \langle a_{(-)}^\dagger a_{(-)}\rangle,
\end{eqnarray}

\noindent
there are nontrivial pairing or polarization functions:

\begin{eqnarray}
\kappa^{(+)}_{n\bar n} &=& \langle e_{n\downarrow} h_{\bar n\uparrow}\rangle, \\
\kappa^{(-)}_{n\bar n} &=& \langle e_{n\uparrow} h_{\bar n\downarrow}\rangle.
\label{eq6}
\end{eqnarray}

In these equations, $n$ and $\bar n$ label electron and hole single-particle states in the dot. $e_n^\dagger$ and $h_{\bar n}^\dagger$ are the corresponding creation operators.
$a_{(+)}$ and $a_{(-)}$ are operators for the two photon polarizations. In the $\kappa$ functions, we have explicitly indicated the electron and hole states that are coupled to a polarized mode. For example, in $\kappa^{(+)}$ the coupling of a spin-down electron state ($m_j=-1/2$) with a ``spin-up'' hole state (coming from a $m_j=-3/2$ electron state in the valence band) is due to the $\sigma_{(+)}$ mode, and it is reinforced by Coulomb interactions.

Notice also that we have assumed pairing between specific electron and hole states,
$n$ and $\bar n$ (orbitals). We will use a harmonic oscillator basis. Thus, for a given
$n$ characterized by radial and angular momentum projection quantum numbers, $n=(k,l)$,
the hole state coupled to $n$ is $\bar n=(k,-l)$. The pair is created in a zero total
angular momentum state, corresponding to the selection rule for interband transitions
\cite{Bastard}.

Mean values, $\langle \dots \rangle$, come from averaging with a density matrix. We
are interested in the very low temperature, $T\to 0$ limit. This density matrix is time
dependent. Thus, the population and polarization functions, Eqs. (\ref{eq1}-\ref{eq6}),
are time-dependent functions. We will derive dynamical equations for their time
evolution (Bloch-Lamb equations).

A further simplification equates $\rho_{nn}^{(e)}=\rho_{\bar n\bar n}^{(h)}=\rho_n$.
This means that the dot is initially neutral, and that we are interested in the dynamics
over a time interval smaller than (or of the order of) the relaxation times in the dot
(the coherent regime \cite{HK}). In fact, we will follow the dynamics in the time
interval (0, 20 ps).

The Hamiltonian describing the quantum dot - microcavity system can be written in the
following form:

\begin{widetext}
\begin{eqnarray}
H &=& \sum_n \left(E_n^{(e)} e_n^\dagger e_n + E_{\bar n}^{(h)}
 h_{\bar n}^\dagger h_{\bar n}\right) +
 \sum_{k,n} \left(t_{kn}^{(e)} e_k^\dagger e_n + t_{\bar k\bar n}^{(h)}
 h_{\bar k}^\dagger h_{\bar n}\right) +\frac{\beta}{2}\sum_{rsuv}
 \langle r,s|\frac{1}{r}|u,v\rangle ~ e_r^\dagger e_s^\dagger
 e_v e_u\nonumber\\
&+&\frac{\beta}{2}\sum_{rsuv}\langle r,s|\frac{1}{r}|u,v\rangle ~
 h_{\bar r}^\dagger h_{\bar s}^\dagger h_{\bar v} h_{\bar u}-\beta\sum_{rsuv}
 \langle r,\bar s|\frac{1}{r}|u,\bar v\rangle ~ e_r^\dagger h_{\bar s}^\dagger
 h_{\bar v} e_u + \hbar\omega \left( a^\dagger_{(+)}a_{(+)}+a^\dagger_{(-)}a_{(-)}\right)
 \nonumber\\
&+& g \sum_n \left\{a^\dagger_{(+)}e_{n\downarrow}h_{\bar n\uparrow}+
 a_{(+)}h^\dagger_{\bar n\uparrow}e^\dagger_{n\downarrow}\right\} +
 g \sum_n \left\{a^\dagger_{(-)}e_{n\uparrow}h_{\bar n\downarrow}+
 a_{(-)}h^\dagger_{\bar n\downarrow}e^\dagger_{n\uparrow}\right\}.
 \label{eq7}
\end{eqnarray}
\end{widetext}

The $E_n$ functions refer to the single-particle energies of the used basis states.
On the other hand, $t_{kn}$ are the matrix elements of the external (quantum dot)
confinement potential. $\beta$ is the strenght of Coulomb interactions, and $g$
the strenght of the electron-photon coupling (assumed constant). The energy of the
photon mode is written as $\hbar\omega$.

\subsection{Bloch-Lamb equations}

The dynamical equations are obtained by taking time derivatives of the
occupation and polarization functions (\ref{eq1}-\ref{eq6}). One gets for
$\rho_n$, for example:

\begin{eqnarray}
i\hbar\frac{{\rm d}\rho_n}{{\rm d}t}=\langle\; [e_n^\dagger e_n,H]\;\rangle.
\end{eqnarray}

\noindent
Once the commutator is computed, we take the mean-field or quasiparticle contribution
(in which mean values of products of four or more operators are expressed in terms of
products of occupations and polarization functions), and neglect the collision terms
\cite{HK}. The result for $\rho_{n\downarrow}$ is:

\begin{eqnarray}
\frac{{\rm d}\rho_{n\downarrow}}{{\rm d}t}=&-&\frac{2\beta}{\hbar}\; {\rm Im}
 \left(\kappa_{n\bar n}^{(+)*} \sum_j \langle n,j|1/r|j,n\rangle
 \kappa_{j\bar j}^{(+)}\right)\nonumber\\
&-&\frac{2 g}{\hbar}\; {\rm Im} \left(\sigma^*_{(+)}\kappa_{n\bar n}^{(+)}\right),
\label{eq9}
\end{eqnarray}

\noindent
and a similar equation for $\rho_{n\uparrow}$. Notice that Coulomb interactions
preserve spin. Thus, if $n$ is a spin-down state (as assumed in Eq. (\ref{eq9})),
the $j$ states entering the sum should be spin-down states also. This is the reason
why only $\kappa^{(+)}$ functions enter the sum.

The equation for the photon number is straightforwardly obtained also, and leads to:

\begin{eqnarray}
\frac{{\rm d}|\sigma_{(+)}|^2}{{\rm d}t}=\frac{2 g}{\hbar}\; {\rm Im}
 \left(\sum_j \sigma_{(+)}^* \kappa_{j\bar j}^{(+)}\right).
\label{eq10}
\end{eqnarray}

Notice the conservation of the polariton number, as it follows from Eqs. (\ref{eq9})
and (\ref{eq10}) (or directly from the Hamiltonian, Eq. (\ref{eq7})): ${\rm d} N_{pol}^{(+)}/{\rm d}t=0$, where:

\begin{eqnarray}
N_{pol}^{(+)}=|\sigma_{(+)}|^2+\sum_n \rho_{n\downarrow}.
\end{eqnarray}

The equations for the polarization functions, $\kappa$, are obtained in the same way.
We shall stress, however, that the modulus, $|\kappa|$, is determined from Eq. (\ref{eq9}),
and it follows that $|\kappa_{n\bar n}^{(+)}|=\sqrt{\rho_{n\downarrow}
(1-\rho_{n\downarrow})}$. Nevertheless, we are forced to use the equation for
$\kappa_{n\bar n}^{(+)}$ in order to determine its phase.

In the same way, the modulus of $\sigma_{(+)}$ (understood as $\langle a_{(+)}\rangle$)
is determined by Eq. (\ref{eq10}), but its phase is not. Thus, we must write the
equation for $\sigma_{(+)}$.

Both $\sigma$ and $\kappa$ should contain a rapidly varying phase factor, $e^{-i\omega t}$,
due simply to the frequency of rotation of the electric field. For example,
$\omega\sim 2\times 10^3$ ps$^{-1}$ for the GaAs band gap. In contrast, we expect
characteristic frequencies of the order of $g/\hbar\sim$ 1 ps$^{-1}$ in the variation of $\rho_n$ or $|\sigma|$. Thus, we will use a kind of Rotating Wave Approximation, and write explicitly the trivial (rapid) phase dependence of $\sigma$ and $\kappa$:

\begin{eqnarray}
\sigma_{(+)}=s_{(+)}\; e^{-i\omega t-i\phi^{(+)}},\label{eq12}\\
\kappa_{n\bar n}^{(+)}=\kappa_{n}^{(+)}\; e^{-i\omega t-i\phi_n^{(+)}},
\label{eq13}
\end{eqnarray}

\noindent
etc. By definition, $s_{(+)}=|\sigma_{(+)}|$, and $\kappa_{n}^{(+)}=
|\kappa_{n\bar n}^{(+)}|$. The equations for $\phi$ and $\phi_n$ are, then:

\begin{eqnarray}
\frac{{\rm d}\phi^{(+)}}{{\rm d}t}=\frac{g}{\hbar s_{(+)}}
 \sum_j \kappa_{j}^{(+)}\cos \left( \phi^{(+)}-\phi_j^{(+)}\right),
\label{eq14}
\end{eqnarray}

\begin{widetext}
\begin{eqnarray}
\frac{{\rm d}\phi_n^{(+)}}{{\rm d}t}&=&-\left\{\omega- \frac{1}{\hbar}
 \left(E_{gap}+E_n^{(e)} + E_{\bar n}^{(h)}+ t_{nn}^{(e)} + t_{\bar n\bar n}^{(h)}\right)
 +\frac{\beta}{\hbar}\langle n,n|1/r|n,n\rangle + \frac{2\beta}{\hbar}
 \sum_{j\ne n} \langle n,j|1/r|j,n\rangle \rho_{j\downarrow}\right\}\nonumber\\
&+& \frac{\beta}{\hbar\kappa_{n}^{(+)}} (2 \rho_{n\downarrow}-1)
 \sum_{j\ne n} \langle n,j|1/r|j,n\rangle\; \kappa_j^{(+)} \cos \left(
 \phi_n^{(+)}-\phi_j^{(+)}\right)
- \frac{g s_{(+)}}{\hbar\kappa_{n}^{(+)}}(2 \rho_{n\downarrow}-1)
 \cos \left( \phi_n^{(+)}-\phi^{(+)}\right),
\label{eq15}
\end{eqnarray}
\end{widetext}

\noindent
and similar equations for $\phi^{(-)}$ and $\phi_n^{(-)}$. We may easily
identify in Eqs. (\ref{eq14},\ref{eq15}) the space-filling factors, $2 \rho_n-1$,
mean-field contributions to the pair energies, detuning factors, etc.

Eqs. (\ref{eq9},\ref{eq15}) are Bloch equations for the electronic occupations and polarizations in the quantum dot \cite{HK}, in a mean field approximation and under a coherent regime, whereas (\ref{eq10},\ref{eq14}) are the analogues of the semiclassical Lamb equations for a laser\cite{QO}. Thus, we will call our system of evolution equations
Bloch-Lamb equations. Losses and incoherent pumping are introduced in the next subsection.

\subsection{Phenomenological account of pumping and losses}
\label{sec2B}

The main processes of the interaction between the quantum dot - microcavity system
and the external medium are described in Ref. \onlinecite{Tejedor}, and modeled
by means of additional terms in the Liouville equation for the density matrix.

One of such processes is the decay of cavity photons by escaping through the
cavity mirrors. We may account for this process by introducing a decay term,
$-\varkappa |\sigma_{(+)}|^2$, into Eq. (\ref{eq10}), with a phenomenological
decay rate, $\varkappa$. Roughly, we have $\varkappa\sim\hbar\omega/Q$, where the
quality factor of the cavity is in the range between 1000 and 5000.\cite{X1}

In terms of the variables introduced in Eqs.
(\ref{eq12},\ref{eq13}), the modified Eq. (\ref{eq10}) is, thus, rewritten:

\begin{eqnarray}
\frac{{\rm d}s_{(+)}}{{\rm d}t}=\frac{g}{\hbar}
 \sum_j \kappa_{j}^{(+)}\sin \left( \phi^{(+)}-\phi_j^{(+)}\right)
 -\varkappa s_{(+)}/2.
\label{eq16}
\end{eqnarray}

On the other hand, the electronic population, $\rho_{n\downarrow}$, is modified
by a continuous incoherent pumping of the quantum dot at a rate $P_{(+)}$, and
the decay of electron-hole pairs by spontaneous emission into leaky modes at a
rate $\gamma$. In Eq. (\ref{eq9}), one should add the following terms:
$P_{(+)} (1-\rho_{n\downarrow})-\gamma \rho_{n\downarrow}$, where Pauli blocking
is explicitly introduced in the pumping contribution. The resulting equation
for $\rho_{n\downarrow}$ is, then:

\begin{widetext}
\begin{eqnarray}
\frac{{\rm d}\rho_{n\downarrow}}{{\rm d}t}=&-&\frac{2\beta}{\hbar}
 \kappa_{n}^{(+)} \sum_{j\ne n} \langle n,j|1/r|j,n\rangle
 \kappa_{j}^{(+)}\sin \left( \phi_n^{(+)}-\phi_j^{(+)}\right)
 -\frac{2 g}{\hbar} s_{(+)}\kappa_{n}^{(+)}\sin \left( \phi^{(+)}-\phi_n^{(+)}
 \right)\nonumber\\
&-& \gamma \rho_{n\downarrow}+P_{(+)} (1-\rho_{n\downarrow}).
\label{eq17}
\end{eqnarray}
\end{widetext}

The set of equations (\ref{eq14}-\ref{eq17}) determine the magnitudes $\rho_{n\downarrow}$,
$s_{(+)}$, $\phi^{(+)}$, and $\phi_n^{(+)}$. $\kappa_{n}^{(+)}$ is obtained in terms
of $\rho_{n\downarrow}$ as $\sqrt{\rho_{n\downarrow}(1-\rho_{n\downarrow})}$.
Notice that the phase of $\kappa_{n\bar n}^{(+)}$ is not affected by the decay
processes in this approximation. The modulus of $\kappa_{n\bar n}^{(+)}$ decays at
a rate which is determined by the decay of $\rho_{n\downarrow}$. This is the analogue
of the relation $T_2=2 T_1$ between the decay times of populations and coherences,
valid both in Atomic Physics \cite{QO} and, in a certain way, in bulk semiconductors
\cite{HK}. Let us stress also that the r.h.s. of Eqs. (\ref{eq14}-\ref{eq17})
depend only on differences of angles. Thus, we may eliminate one of the angles, for
example $\phi$, by introducing the new variables $\theta_n=\phi-\phi_n$.

A formally similar set of equations may be written for $\rho_{n\uparrow}$, $s_{(-)}$, $\phi^{(-)}$, and $\phi_n^{(-)}$. Notice that, in our approximation, where Coulomb
collision terms and other relaxation processes (due to the interaction with phonons,
impurities, etc) are neglected, the relations $\rho_{nn}^{(e)}=\rho_{\bar n\bar n}^{(h)}$
are preserved at all times, and $\rho_{n\downarrow}$ evolves independently of
$\rho_{n\uparrow}$.

\subsection{The $t \to 0$ asymptotics}
\label{sec2c}

Eqs. (\ref{eq14}-\ref{eq17}) are to be solved with null initial conditions, i.e., there
are no pairs in the quantum dot and no photons in the cavity at $t=0$.

According to Eq. (\ref{eq17}), the populations $\rho_{n\downarrow}$ rise as
$P_{(+)} t$ and, consequently, $\kappa_{n}^{(+)}\sim \sqrt{P_{(+)} t}$. However,
the small-$t$ asymptotics of the magnitudes $s_{(+)}$, $\phi^{(+)}$ and $\phi_n^{(+)}$
should be analysed with care because of terms like $\kappa_{n}^{(+)}/s_{(+)}$, etc
entering the equations. We found that the consistent behavior in the $t\to 0$ limit
is the following:

\begin{eqnarray}
\rho_{n\downarrow}&=&P_{(+)} t + \dots, \label{eq18}\\
\kappa_{n}^{(+)}&=&\sqrt{P_{(+)} t}+\dots, \\
s_{(+)}&=&\frac{2 g N_{states}}{3\hbar}\sqrt{P_{(+)}}\; t^{3/2}+\dots, \\
\phi_{n}^{(+)}&=&a_n^{(+)} t+\dots, \\
\phi^{(+)}&=&\pi/2+a_{\phi}^{(+)} t+\dots,
\end{eqnarray}

\noindent
where $N_{states}$ is the number of single-particle states available in the dot, and

\begin{eqnarray}
a_n^{(+)}&=&-\omega + \frac{1}{\hbar}
 \left(E_{gap}+E_n^{(e)}+E_{\bar n}^{(h)}+t_{nn}^{(e)}+t_{\bar n\bar n}^{(h)}\right)
 \nonumber\\
&-&\frac{\beta}{\hbar} \sum_j \langle n,j|1/r|j,n\rangle, \label{eq23}\\
a_\phi^{(+)}&=&\frac{3}{5 N_{states}} \sum_n a_n^{(+)}.
\label{eq24}
\end{eqnarray}

\noindent
$n$ and $j$ in Eqs. (\ref{eq23}-\ref{eq24}) represent spin-down states.
Eqs. (\ref{eq18}-\ref{eq24}) are to be used in the numerical integration of Eqs.
(\ref{eq14}-\ref{eq17}).

\section{Numerical results}
\label{sec3}

In order to perform the numerical integration of Eqs. (\ref{eq14}-\ref{eq17}),
we shall give a precise meaning to the magnitudes entering them. We are interested
in a quantum dot which supports a finite number of single-particle states,
$N_{states}$. Thus, $N_{states}$ will be fixed by hand. A magnetic field, $B=7$ Teslas, acts on the charge carriers in the dot. Both, the
dot confinement and the magnetic field may be thought of as control parameters.
Energy magnitudes are converted to angular frequencies by dividing by $\hbar$.

We will use parameters for GaAs. The single-particle energies are defined as:

\begin{eqnarray}
E_n^{(e,h)}&=&\frac{\hbar\omega_c^{(e,h)}}{2}
 \left( 2 k_n \pm l_n + |l_n| +1 \right)\nonumber\\
&\pm& g_{e,h} \mu_B B S_{zn}^{(e,h)},
\label{eq25}
\end{eqnarray}

\noindent
where the cyclotron frequencies are $\omega_c^{(e)}=2.625\; B$ ps$^{-1}$,
$\omega_c^{(h)}=1.173\; B$ ps$^{-1}$ (and $B$ in Teslas). $S_z=\pm 1/2$ are
the spin projections along $B$. $g_e\mu_B/(2\hbar)=4.397\times 10^{-3}$ ps$^{-1}$,
and $g_h\mu_B/(2\hbar)=-4.397\times 10^{-2}$ ps$^{-1}$. The + sign in Eq.
(\ref{eq25}) corresponds to the electron. The fact that, in our simple model, energy levels come from a parabolic confinement and a magnetic field has not a decisive importance. What really matters is the position of the bare levels, and the effects of Coulomb interaction
among carriers, as will become evident in the next sections.

On the other hand, the matrix elements of the (parabolic) confinement potential are written as:

\begin{eqnarray}
t_{nn}^{(e,h)}&=&\frac{\hbar\omega_0^2}{\omega_c^{(e,h)}}
 \left( 2 k_n + |l_n| +1 \right),
\end{eqnarray}

\noindent
where the dot frequency is $\omega_0=4.558$ ps$^{-1}$. The electron-photon coupling,
$g$, which determines the Rabi frequency,
is fixed to $g/\hbar=1.5$ ps$^{-1}$, and the strength of Coulomb interactions is
$\beta/\hbar=3.871 \sqrt{B}$ ps$^{-1}$.

Notice that we have not included the nominal gap energy into the electron energy, Eq.
(\ref{eq25}). A shifted mode frequency is defined according to $\Delta=\omega-E_{gap}/\hbar$.

Concerning the loss and pumping parameters introduced in Sec. \ref{sec2B}, we will set
the rate of spontaneous decay, $\gamma$, to 0.1 ps$^{-1}$. The equations are
integrated for differents sets of $P$, $\varkappa$, and $\Delta$. A variation of
$\Delta$ should be understood as varying the cavity dimensions. In fact, $B$,
$\omega_0$ and $P$ are the only parameters which can be externally controlled for a given
dot-microcavity system.

Once we have precisely defined our equations, we may turn to the analysis of the
simplest system, in which the dot may support only one pair.

\begin{figure}[t]
\begin{center}
\includegraphics[width=.95\linewidth,angle=0]{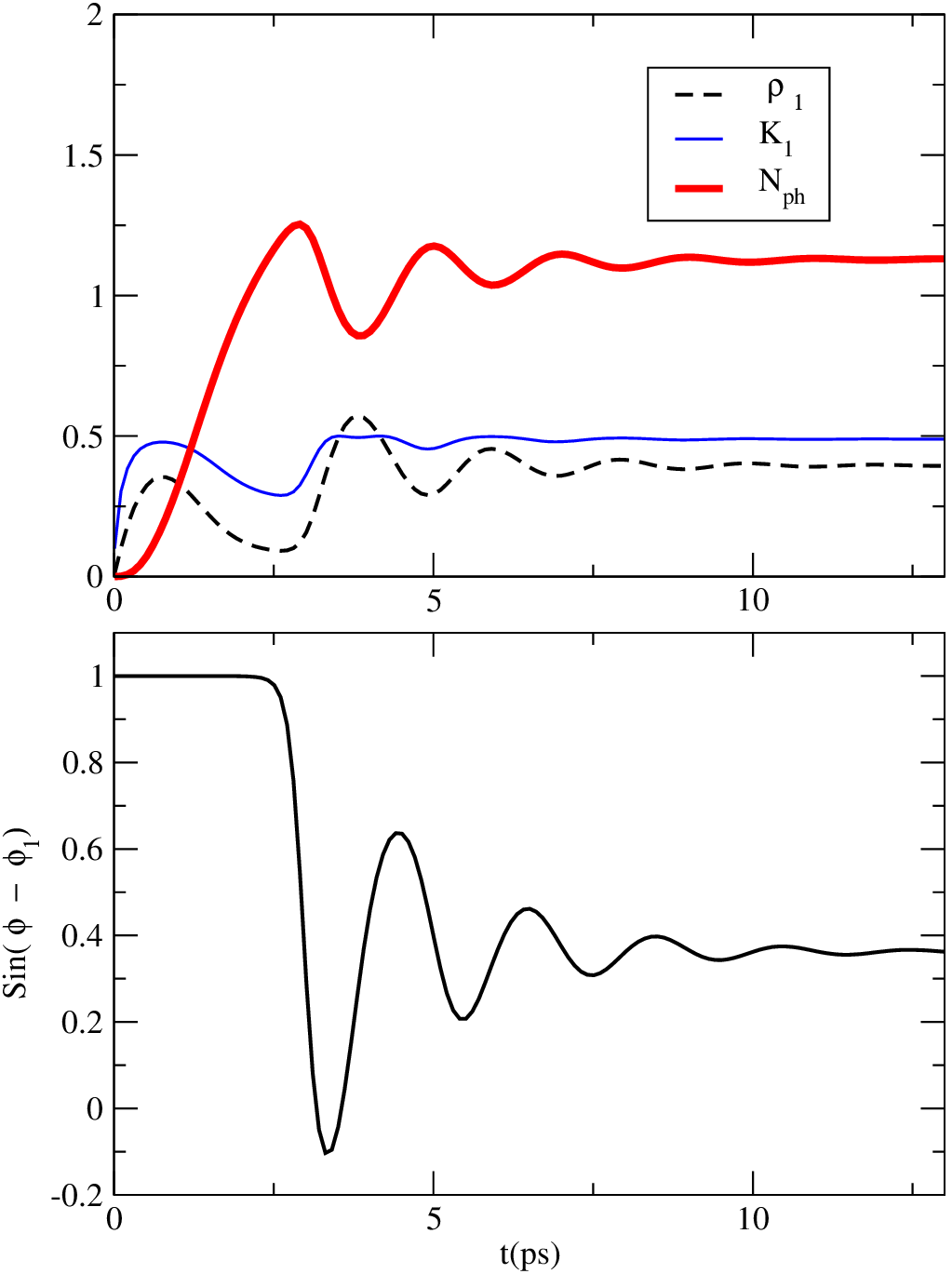}
\caption{\label{fig1} (Color online) Number of photons, dot occupation and polarization
 in the one-pair dot for $P=1$ ps$^{-1}$, $\varkappa=0.5$ ps$^{-1}$. The photon energy is
 $\Delta=4.5$ ps$^{-1}$ (maximum resonance). }
\end{center}
\end{figure}

\subsection{A shallow dot with $N_{states}=1$}

This case was extensively studied in Ref. [\onlinecite{Tejedor}]. The authors obtain
and solve the equations for the density matrix in a Fock basis composed by the
electron-hole vacuum, $|G\rangle$, the exciton ground state, $|X\rangle$, and the
$n$-photon states, $|n\rangle$. They write equations for the time evolution of the
populations, $\rho_{Gn,Gn}$, $\rho_{Xn,Xn}$, and the coherences, $\rho_{Gn,Xn-1}$.
To solve them, they should truncate the photon Hilbert space. For $n\le 100$, for
example, they have 300 equations. This is an exact treatment, in which the authors may compute correlation functions for the photon field, the spectrum of the emitted light, etc.

In our model, however, the $N_{states}=1$ case is described by only three variables,
$\rho_1$, $s$, and the phase difference $\phi-\phi_1$. The photon field is assumed
coherent. In spite of these simplifications, most of the results presented in
paper [\onlinecite{Tejedor}] concerning the system dynamics are qualitatively and quantitatively reproduced by our equations. In the present section, we focus on new results, uncovered in paper [\onlinecite{Tejedor}].

The first feature we observe in our calculations is the presence of the
three regimes, mentioned in [\onlinecite{Tejedor}], under which the quantum dot -
microcavity system may operate. They can be roughly classified according to the
relation between pumping and losses: (i) $P\gg \varkappa$, (ii) $P\sim\varkappa$,
and (iii) $\varkappa\gg P$.

We show in Fig. \ref{fig1} an example of calculations corresponding to the intermediate regime (ii), in which $P=1$ ps$^{-1}$ and $\varkappa=0.5$ ps$^{-1}$. This example reveals oscillations in the transient interval until the asymptotic values are reached.
The oscillations of the number of photons in the cavity are due to non compensation between the rates of creation of photons and losses, and even to photon absorption by the dot (when $\sin (\phi-\phi_1) < 0$).

\begin{figure}[t]
\begin{center}
\includegraphics[width=.95\linewidth,angle=0]{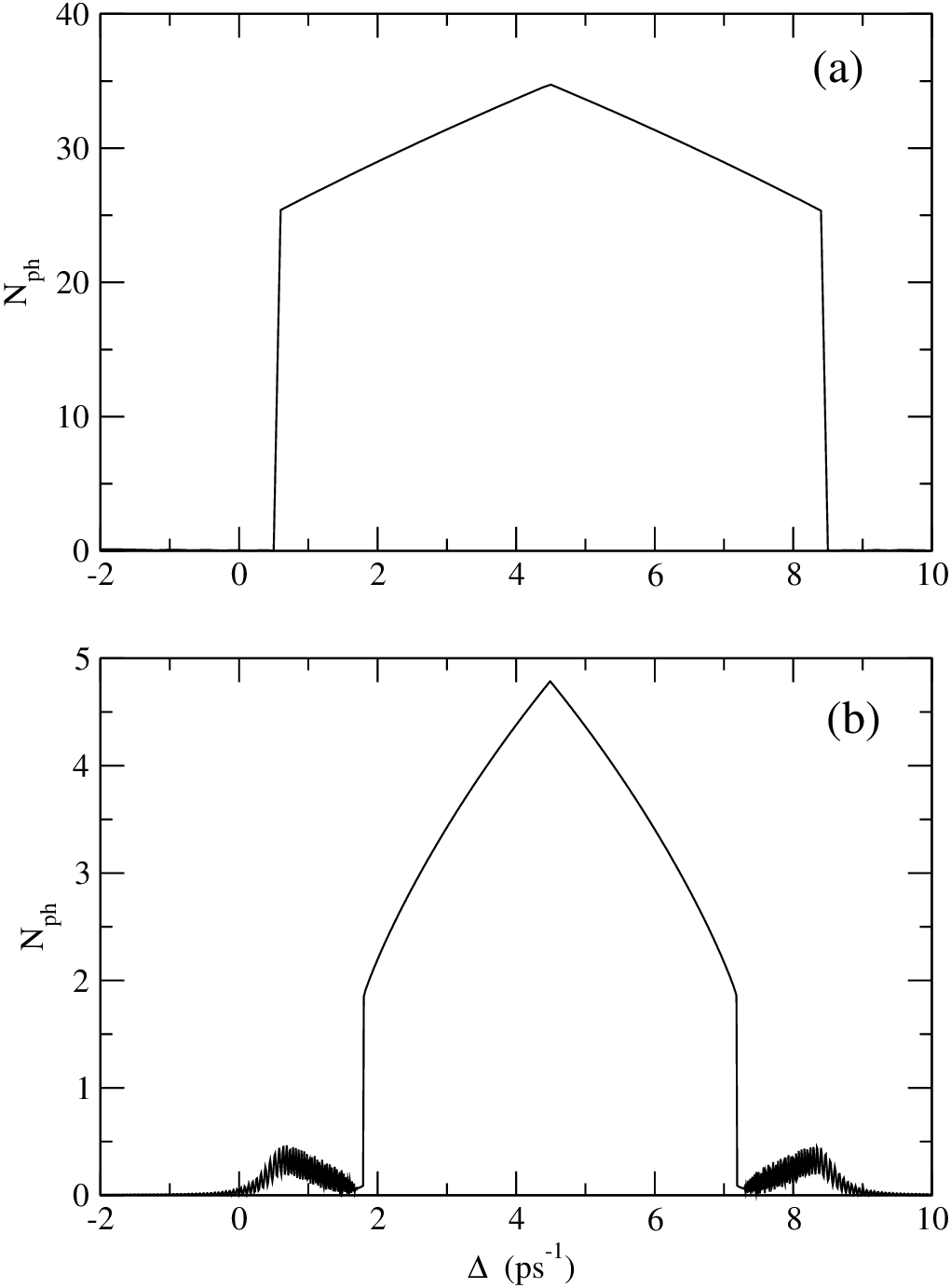}
\caption{\label{fig2} The number of photons as a function of the photon energy for
 (a) $P=7$ ps$^{-1}$, $\varkappa=0.1$ ps$^{-1}$, and (b) $P=1$ ps$^{-1}$, $\varkappa=0.1$
 ps$^{-1}$.}
\end{center}
\end{figure}

The asymptotic number of photons, computed at $t_f=100$ ps as a function of $\Delta$ for a high quality cavity ($\varkappa=0.1$) is shown in Fig. \ref{fig2}. The pair frequency (pair energy divided by $\hbar$) in the present case is

\begin{eqnarray}
\omega_{pair}&=& \frac{1}{\hbar} \left(E_1^{(e)} + E_{\bar 1}^{(h)}+ t_{11}^{(e)} +
 t_{\bar 1\bar 1}^{(h)}\right)\nonumber\\
 &-&\frac{\beta}{\hbar}\langle 1,1|1/r|1,1\rangle=4.5\; {\rm ps}^{-1}.
\end{eqnarray}

\noindent
Extreme resonance corresponds to $\Delta=\omega_{pair}$. The dependence $N_{ph}$ vs
$\Delta$ in Fig. \ref{fig2}a ($P=7$ ps$^{-1}$) resembles a step function with a mean value of 30 photons inside the resonance interval, and zero outside. If the cavity is such that the actual $\Delta$ lies in the abruptly varying region, small changes produced by variations in $B$ or in $\omega_0$ could switch between the two ``states'' of the cavity. Although the step-like variation could be an artifact of our simplified equations, an abrupt variation is expected.

\begin{figure}[t]
\begin{center}
\includegraphics[width=.95\linewidth,angle=0]{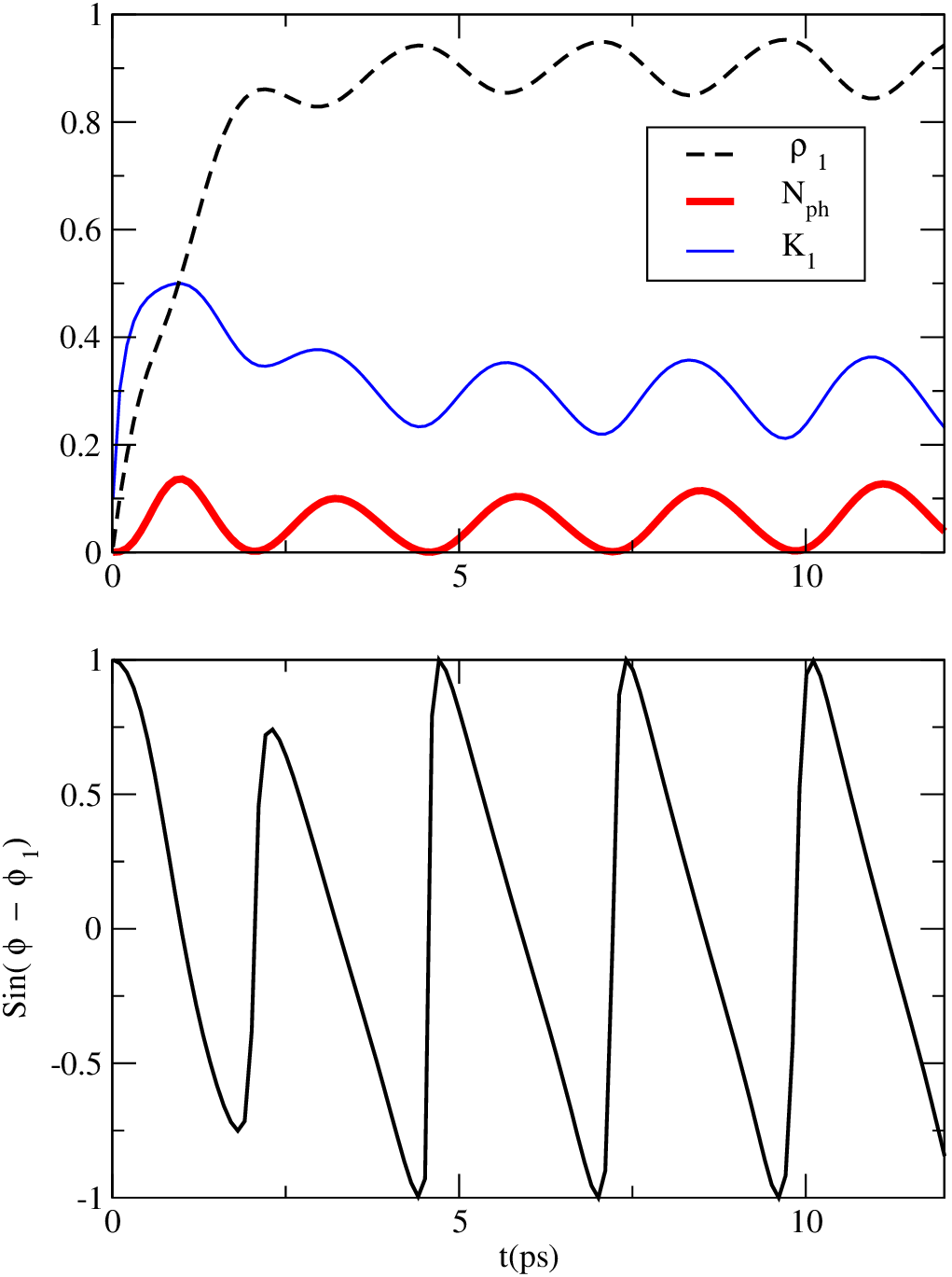}
\caption{\label{fig3} (Color online) The same as in Fig. \ref{fig1}, but for
 $P=1$ ps$^{-1}$, $\varkappa=0.1$ ps$^{-1}$, and $\Delta=8$ ps$^{-1}$.}
\end{center}
\end{figure}

Similar step-like variations are shown in Fig. \ref{fig2}b, which corresponds to $P=1$ ps$^{-1}$. In addition, outside the resonance interval there are asymptotic oscillating solutions, whose characteristic behavior is illustrated in Fig. \ref{fig3} for $\Delta=8$ ps$^{-1}$. Notice that, in the oscillating solution, the quantum dot is periodically emitting ($\sin (\phi-\phi_1) > 0$) and absorbing ($\sin (\phi-\phi_1) < 0$) photons from the cavity with a period around 2.5 ps. The mean number of photons is only around 0.1, thus our coherent approximation could be very rough. Such periodic solutions are, however, very common in the larger systems with $N_{states}>1$, as will be seen below.

\begin{figure}[t]
\begin{center}
\includegraphics[width=.95\linewidth,angle=0]{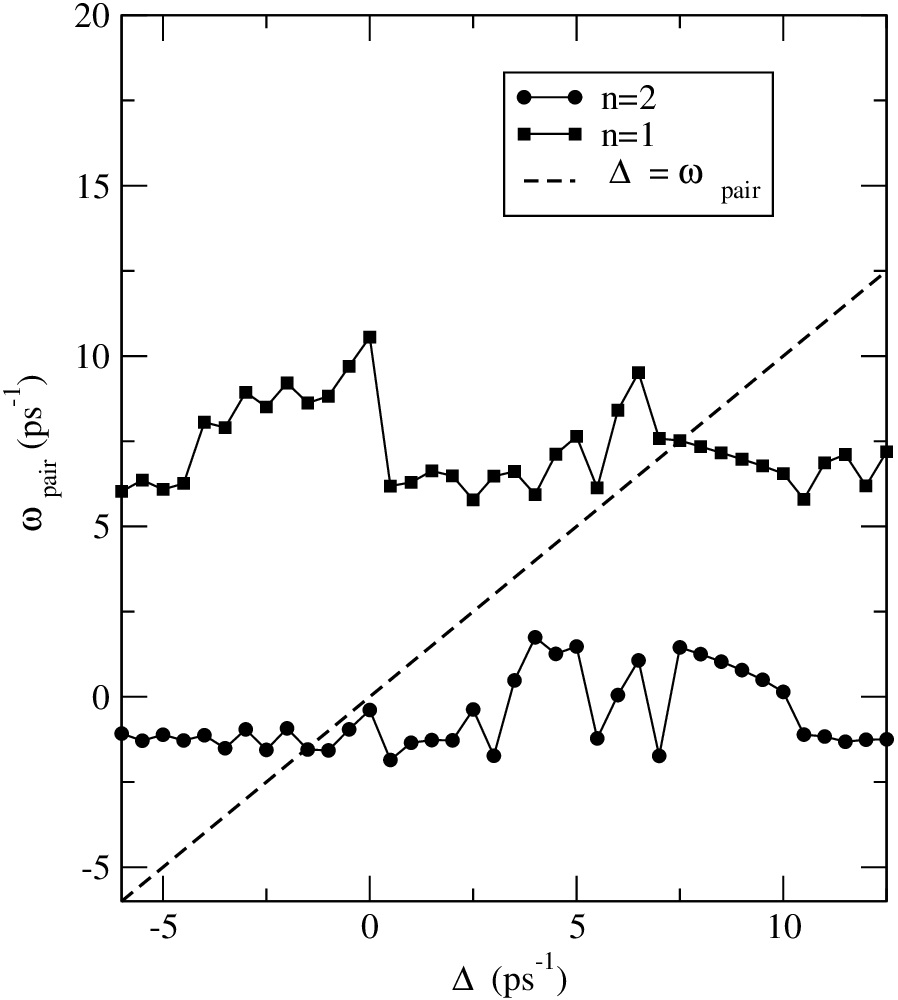}
\caption{\label{fig4} The single-pair levels as a function of the photon energy for
 $P=1$ ps$^{-1}$, $\varkappa=0.1$ ps$^{-1}$.}
\end{center}
\end{figure}

From the point of view of the theory of nonlinear evolution differential equations\cite{chaos}, in the resonance interval the solution of the system (\ref{eq14}-\ref{eq17}) is attracted by a stable critical point (an spiral), whereas the asymptotic periodic solution outside the resonace interval for $\Delta$ corresponds to a limit cycle. This view reveals the existence of a second stable attractor (also an spiral) in the resonance interval, which is not seen in calculations because the used initial conditions is not in its basin of atraction. As we move out of the resonance interval, one of the stable critical points dissapears (a bifurcation) and the limit cycle emerges. In addtition, at any value of $\Delta$ there is a third critical point of hyperbolic character (with one unstable direction). The existence of two asymptotically stable states of the quantum dot-microcavity system, how to reach them, and even how to reach the hyperbolic point by using control of chaos \cite{chaos}, the nature of the limit cycles, etc are very interesting questions which deserve further investigation. In the larger, $N_{states}>1$, systems the dynamics is expected to be even more complex. In the present paper, we restrict the analysis to the situation with null initial conditions, described in Sec. \ref{sec2c}, and leave the more general questions for a later work.

\subsection{The $N_{states}=2$ dot coupled to $\sigma_{(+)}$ light}

In view of the fact that, in our equations, the $\sigma_{(+)}$-polarized mode evolves independently of the $\sigma_{(-)}$ mode, the next nontrivial case corresponds to a dot which may support up to two pairs, $N_{states}=2$, interacting with the same photon mode. To be specific, we will consider coupling to the $\sigma_{(+)}$ mode.

The dynamics, in many ways, has features very similar to those of the simpler one-pair
system. In particular, the three operation regimes are observed. In addition, there
are new characteristics, which are briefly outlined below.

\begin{figure}[t]
\begin{center}
\includegraphics[width=.95\linewidth,angle=0]{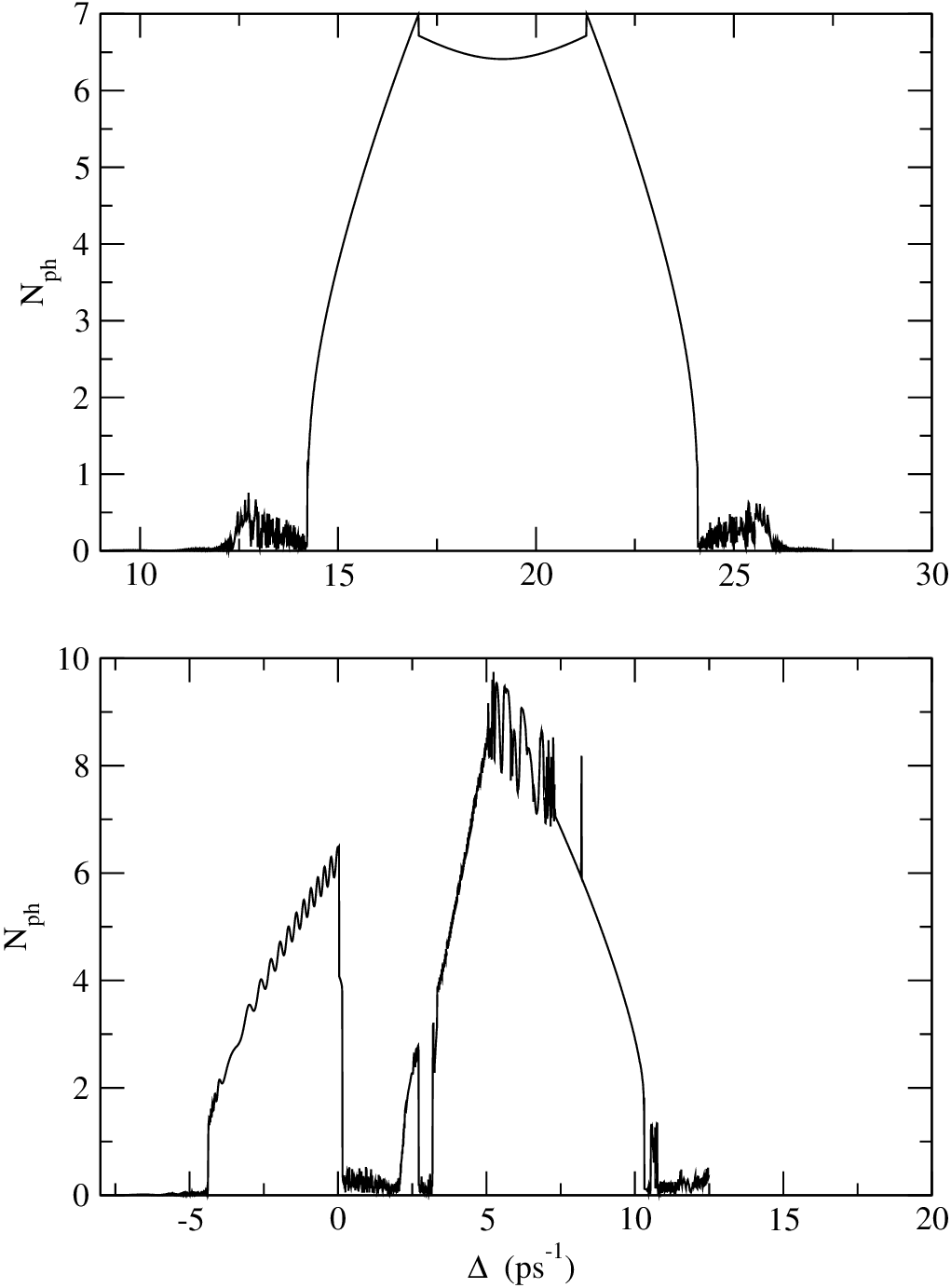}
\caption{\label{fig5} The number of photons as a function of the photon energy for the $N_{states}=2$ system with $P=1$ ps$^{-1}$, $\varkappa=0.1$ ps$^{-1}$. In the upper
panel, Coulomb interactions are not included ($\beta=0$).}
\end{center}
\end{figure}

The first new (expected) feature is the two-peak structure in the number of
photons as a function of $\Delta$ as a result of individual resonances with the
single-pair levels. Coulomb interactions are crucial in determining the position
of these levels. Indeed, we have:

\begin{eqnarray}
\omega_{pair}(n)&=&\frac{1}{\hbar} \left(E_{gap}+E_n^{(e)} + E_{\bar n}^{(h)}+
 t_{nn}^{(e)} + t_{\bar n\bar n}^{(h)}\right)\nonumber\\
 &-&\frac{\beta}{\hbar}\langle n,n|1/r|n,n\rangle\nonumber\\
 &-& \frac{2\beta}{\hbar} \sum_{j\ne n} \langle n,j|1/r|j,n\rangle \rho_{j\downarrow}.
\end{eqnarray}

In our model quantum dot, the pair frequencies at $\beta=0$ are igual to 17.32 and
20.99 ps$^{-1}$. The account of Coulomb interactions move them to the interval from
-2 to 10 ps$^{-1}$, as shown in Fig. \ref{fig4}, and make the single-pair levels
dependent on the occupations $\rho_n$. But Coulomb interactions modify also the phases
and occupations (see Eqs. (\ref{eq15}) and (\ref{eq17})) causing strong changes in
the stationary solutions. We illustrate in Fig. \ref{fig5} these effects for the
system with parameters $P=1$ ps$^{-1}$ and $\kappa=0.1$ ps$^{-1}$.

\begin{figure}[t]
\begin{center}
\includegraphics[width=.95\linewidth,angle=0]{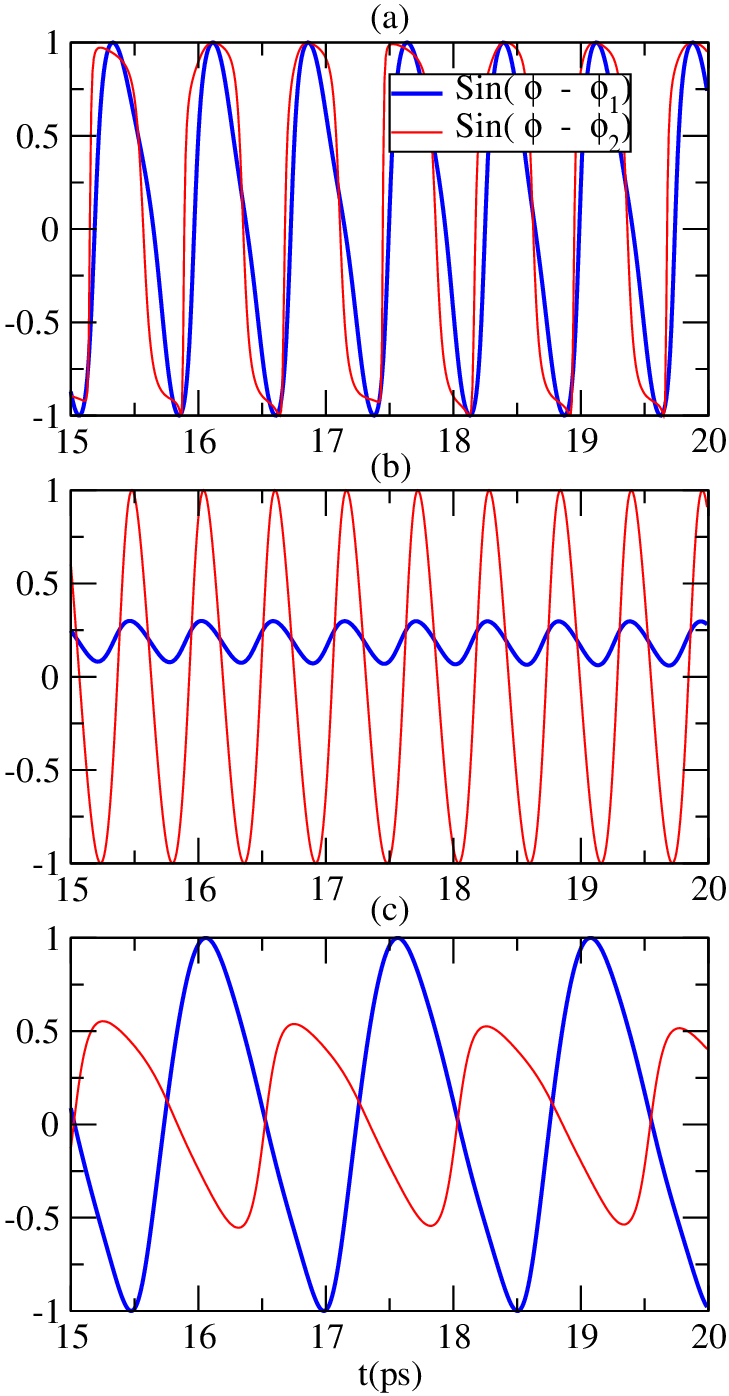}
\caption{\label{fig6} (Color online) $\sin (\phi-\phi_i)$
in the $N_{states}=2$ dot and cavity parameters $P=1$ ps$^{-1}$, $\varkappa=0.1$ ps$^{-1}$,
and (a) $\Delta=7$ ps$^{-1}$, (b) $\Delta=-3$ ps$^{-1}$, and (c) $\Delta=2.5$ ps$^{-1}$.}
\end{center}
\end{figure}

A second interesting characteristic is that asymptotic oscillatory solutions are very
common, in particular inside the resonance intervals. It means that, as a rule, the
number of photons and the occupations will vary periodically around a mean value. In our Fig. \ref{fig5}, where the number of photons is computed at fixed $t_f=100$ ps, oscillatory solutions are seen as apparently random behavior of $N_{ph}$.

It is interesting also to look at the relative phases, $\phi-\phi_i$ in each of the
resonance intervals. Let us consider, for example, the system described in Fig.
\ref{fig5}b. In the upper interval, $3\lesssim\Delta\lesssim 10$ ps$^{-1}$, the two
single-pair levels emit or absorb almost in phase (see Fig. \ref{fig6}a). In the
lowest interval, $-5\lesssim\Delta\lesssim 0$ ps$^{-1}$, one level is emitting, and the
second one emits and absorb, but they do so ``in phase''. That is, maximum emission
occurs at the same time in both levels, and maximum absorption in one of them occurs
at coincidence with minimum emission in the other. The situation is depicted in Fig.
\ref{fig6}b. Finally, in the central peak around $\Delta\approx 2.5$ ps$^{-1}$, which is
the result of interference and dissapears when parameters are changed,
emission and absorption in the two levels are ``in counterphase'', as can be seen in
Fig. \ref{fig6}c. Notice that the periods of these oscillatory motions take values in the range between 0.5 and 1.5 ps. This means angular frequencies between 4 and 12 ps$^{-1}$.
These numbers may be compared with the Rabi frequency, $g\;s/\hbar=1.5\;s$ ps$^{-1}$, which takes values around 2 - 4 ps$^{-1}$, and the characteristic frequency of Coulomb
interactions, $\beta/\hbar\approx 10$ ps$^{-1}$.

Let us stress that we are pumping both single-pair levels at the same rate, $P$. The
qualitative features of the $N_{states}=2$ problem, mentioned above, do not seem to
rely on the specific pumping scheme used. The qualitative features dot not change either
when relaxation from the higher-energy level to the lower-energy one is included.

\begin{figure}[t]
\begin{center}
\includegraphics[width=.95\linewidth,angle=0]{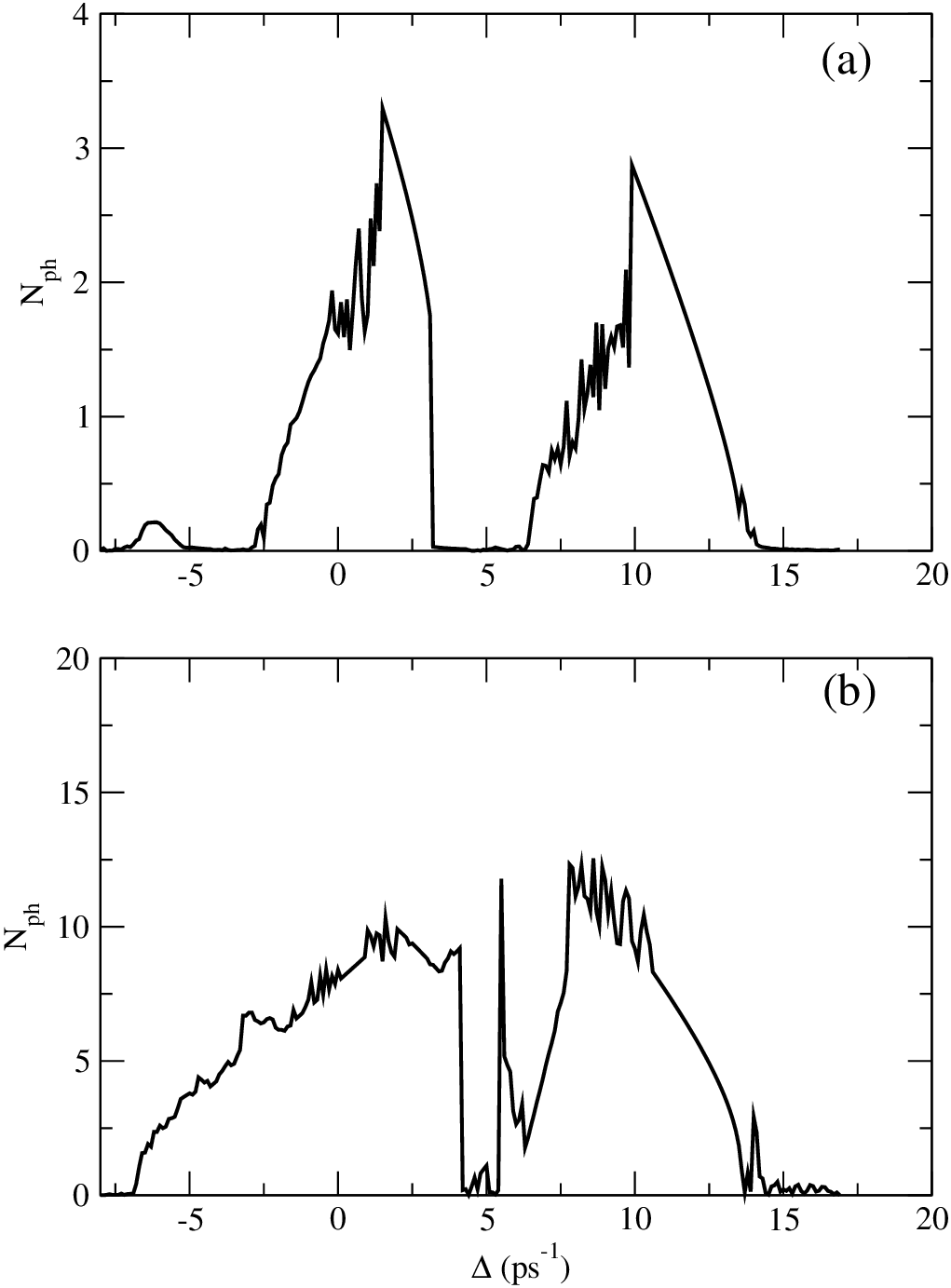}
\caption{\label{fig7} Number of photons as a function of the photon energy for the
$N_{states}=3$ dot and cavity parameters (a) $P=3$ ps$^{-1}$, $\varkappa=1$ ps$^{-1}$,
and (b) $P=1$ ps$^{-1}$, $\varkappa=0.1$ ps$^{-1}$.}
\end{center}
\end{figure}

\subsection{The $N_{states}=3$ dot coupled to $\sigma_{(+)}$ light}

In general, the resonance curve, $N_{ph}$ vs $\Delta$, in the $N_{states}=3$ system
exhibits three peaks, as one can expect (Fig. \ref{fig7}a). For certain values of the
system parameters, two or more peaks may overlap, or additional sharp peaks (the result of interference) may emerge (see, for example, Fig. \ref{fig7}b).

Concerning the relative phases between the polarization functions and the radiation field,
$\phi-\phi_i$, one can roughly say the following. In the uppermost resonance interval, the three single-pair levels cooperate. This means that they are all emitting simultaneously,
or periodically absorbing-emitting almost ``in phase''. In the central interval, in
general, there are two levels cooperating, and the third is not. For example, two of
them are emitting, and the third is absorbing. Finally, in the lowest interval, when one
of the levels is emitting, the second is absorbing, and the third alternates between the
first two. Curves are qualitatively similar to the $N_{states}=2$ case, and will not be
shown.

\begin{figure}[t]
\begin{center}
\includegraphics[width=.95\linewidth,angle=0]{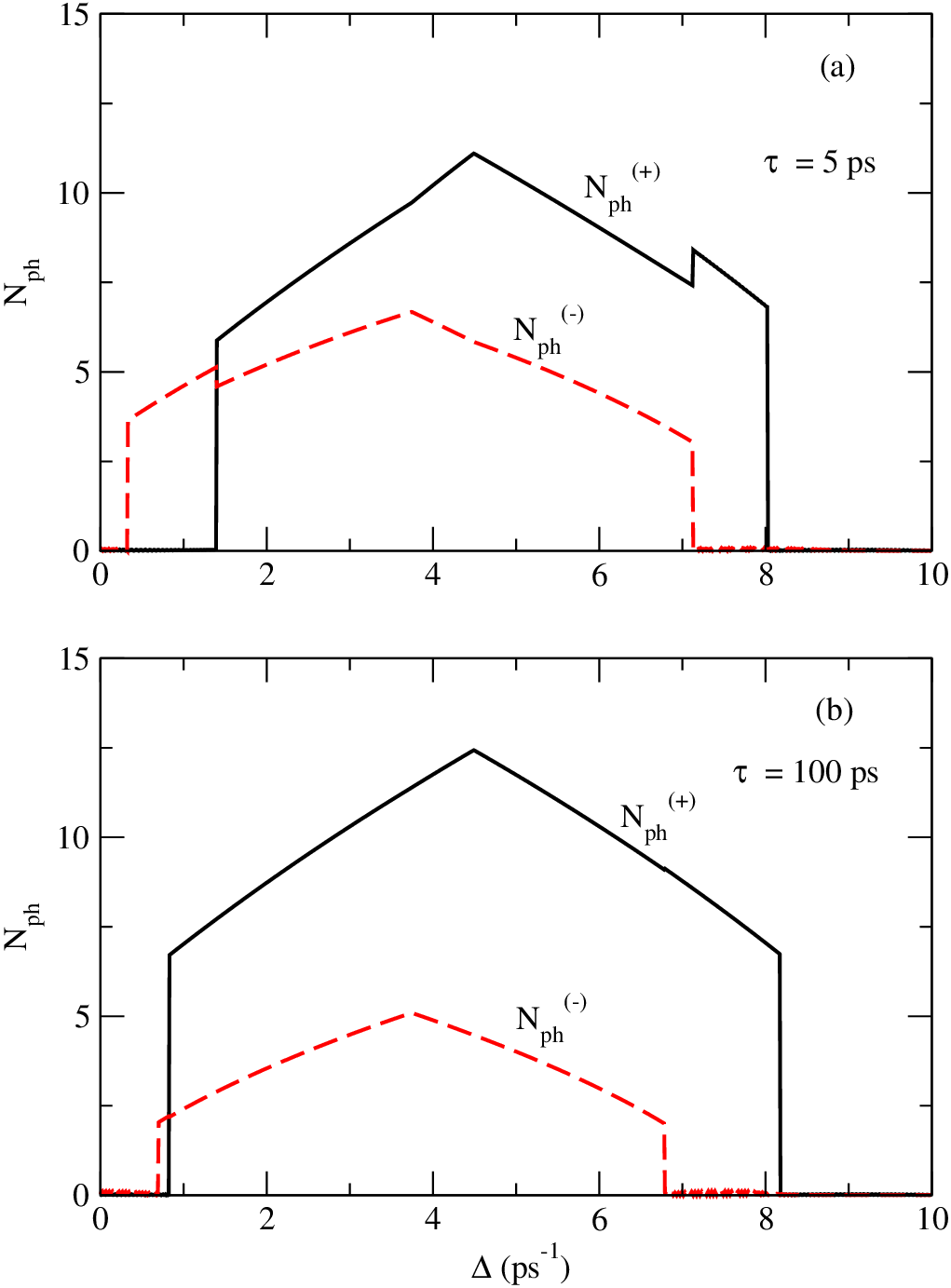}
\caption{\label{fig8} (Color online) The $N_{states}=2$ dot with spin-flip
processes included: (a) $\tau_s=5$ ps, (b) $\tau_s=100$ ps. See explanation
in the main text.}
\end{center}
\end{figure}

\subsection{The $N_{states}=2$ dot coupled to different photon modes}

In this section, we consider the possibility of spin-flips between exciton states with a
characteristic rate $2\pi/\tau_s$. Typical values of $\tau_s$ are around 5 ps or
larger \cite{taus}. Then, there is a new nontrivial situation for the $N_{states}=2$
dot, in which one electron-hole pair is coupled to the $\sigma_{(+)}$ photon mode,
and the second one to the $\sigma_{(-)}$ mode.

The spin-flip term will be added to the equations for the occupations, Eq. (\ref{eq17}).
At $B=7$ Teslas, the Zeeman splitting between exciton states is around 0.7 meV. The
lowest exciton state is coupled to the $\sigma_{(-)}$ photon. Then, in the equation for
$\rho_{1\downarrow}$ (the exciton coupled to the $\sigma_{(+)}$ photon), we add a term

\begin{equation}
-\frac{2\pi}{\tau_s} \rho_{1\downarrow} (1-\rho_{1\uparrow}),
\end{equation}

\noindent
and for $\rho_{1\uparrow}$ a term

\begin{equation}
\frac{2\pi}{\tau_s} \rho_{1\downarrow} (1-\rho_{1\uparrow}).
\end{equation}

\noindent
Notice the Pauli blocking factors in them.
Spin-flip transitions from $\rho_{1\uparrow}$ to $\rho_{1\downarrow}$ would require an
activation mechanism to overcome the energy barrier, and will not be included.

The results for the system with cavity parameters $\varkappa=0.1$ ps$^{-1}$, $P_{(+)}=5$ ps$^{-1}$, $P_{(-)}=2$ ps$^{-1}$, are shown in
Fig. \ref{fig8}. The lower figure corresponds to $\tau_s=100$ ps, i.e., when the
$\sigma_{(+)}$ and $\sigma_{(-)}$ dynamics are independent. The upper one, to
$\tau_s=5$ ps, that is, a rate $2\pi/\tau_s\approx 1.2$ ps$^{-1}$. Step-like variations in one of the photon populations are transmitted to the other photon number. Interesting
enough is the region $0.4 < \Delta < 1.3$ ps$^{-1}$, where there are only $\sigma_{(-)}$
photons in spite of the fact that $P_{(+)}>P_{(-)}$.

\section{Concluding remarks}
\label{sec4}

We have explicitly written and numerically solved the evolution (Bloch-Lamb) equations
describing the dynamics of a quantum dot-microcavity system. The main simplifications
contained in our equations are the following: (a) they are mean field equations, i.e.,
dot not contain particle-particle correlations, and (b) the photon field is coherent.
Assumption (b) is closely related to (a). The reward from this simplified approach is
that the system of equations is relatively small, allowing a qualitative analysis of
its solutions and the extension to larger dots, which may support more than one
electron-hole pair.

In the simplest system with $N_{states}=1$, the qualitative analysis revealed the
existence of resonance intervals for the parameters, where the coupling to the photon
field of the cavity is optimal and the number of photons in the stationary state
reaches a maximum. There is also a second stable stationary state, which is not reached
because the used (null) initial conditions is not in its basin of attraction. The two stable solutions should be further analysed focusing on a possible optical bistability
in this system\cite{bistability}. Outside the resonance interval, one of the stable states dissapears (suggesting a bifurcation) leading to the appearence of stable periodic orbits in which the physical magnitudes oscillate. In this oscillatory states, the quantum dot is periodically absorbing and emitting photons from and to the cavity. Under strong pumping,
the transition resonance-out of resonance may be almost step-like, a fact which may also
have interesting applications. We have left many questions unexplored such as, for
example, the width of the resonance interval in $\Delta$, or the transient time before the
stationary state is reached, as functions of the system parameters.

Let us stress that the existence of two stable stationary solutions are intuitively related to the two possible couplings between an exciton and a photon (the two polariton branches in a quantum well under weak coupling, for example). However, we can not yet understand the way they may arise from the system of linear equations of Ref. \onlinecite{Tejedor}. We think that the truncation fo the photon Fock space is responsible for this difference with our results.

Larger systems exhibit even more complex dynamics, which seems to strongly depend on the
position of the single-pair levels. Most of the new features are already seen in the
$N_{states}=2$ dot, when both pair states are coupled to the same photon mode. We showed
results for a dot in which the energy levels are 8 - 10 meV appart, in such a way that
individual resonance intervals are resolved. Coulomb interactions are shown to have
drastic effects, not only on the level positions, but also causing interferences and
altering the very nature of the stationary states. Oscillatory states are reached almost
for any values of the system parameters, even inside the resonance intervals. Cooperative
or non cooperative behavior of the single-pair levels are obtained in different parameter
regions. In the $N_{states}=3$ system, these features are reinforced.

The account of relaxation between electronic states or a change in the pumping scheme
used to feed the dot are shown to have small impact on this qualitative and
semiquantitative picture. Other effects such as the coupling to the delocalized states
of the wetting layer surrounding the self-assembled dot are to be considered. This
coupling to the wetting layer is crucial in other contexts \cite{wl}.

Work along some of the abovementioned questions is currently in progress.

\begin{acknowledgments}
The authors acknowledge the Comitee for Research of the Universidad
de Antioquia for support.
\end{acknowledgments}


\begin{thebibliography}{99}
\bibitem{micromaser} D. Meschede, H. Walther, and G. Muller, Phys. Rev.
 Lett. {\bf 54}, 551 (1985).
\bibitem{Tejedor} J.I. Perea, D. Porras, and C. Tejedor,
 http://arxiv.org/cond-mat/0310570.
\bibitem{X1} J.M. Gerard, B. Sermage, B. Gayral, B. Legrand, E. Costard,
 and V. Thierry-Mieg, Phys. Rev. Lett. {\bf 81}, 1110 (1998).
\bibitem{X2} G.S. Solomon, M. Pelton, and Y. Yamamoto, Phys. Rev. Lett.
 {\bf 86}, 3903 (2001).
\bibitem{X3} M. Pelton, C. Santori, J. Vuckovic, B. Zhang, G.S. Solomon,
 J. Plant, and Y. Yamamoto, Phys. Rev. Lett. {\bf 89}, 233602 (2002).
\bibitem{XX1} O. Benson, C. Santori, M. Pelton, and Y. Yamamoto, Phys. Rev. Lett.
 {\bf 84}, 2513 (2000).
\bibitem{XX2} T.M. Stace, G.J. Milburn, and C.H.W. Barnes, Phys. Rev. {\bf B 67},
 085317 (2003).
\bibitem{radius} J.M. Gerard, D. Barrier, J.Y. Marzin, R. Kuszelewicz, L. Manin,
 E. Costard, V. Thierry-Mieg, and T. Rivera, Appl. Phys. Lett. {\bf 69}, 449
 (1996).
\bibitem{modos} A. Yariv, {Optical Electronics} (Saunders College,
 San Francisco, 1991).
\bibitem{Bastard} G. Bastard, {\it Wave mechanics applied to
 semiconductor heterostructures} (Les editions de physique, Les Ulis
 Cedex, 1998).
\bibitem{HK} H. Haug and S.W. Koch, {\it Quantum Theory of the Optical
 and Electronic Properties of Semiconductors} (World Scientific,
 Singapore, 1994).
\bibitem{QO} M.O. Scully and S. Zubairy, {\it Quantum Optics}
 (Cambridge Univ. Press, Cambridge, 2001).
\bibitem{chaos} E. Ott, {\it Chaos in Dynamical Systems} (Cambridge
 Univ. Press, Cambridge, 2002).
\bibitem{taus} J.M. Kikkawa and D.D. Awschalom, Phys. Rev. Lett.
 {\bf 80}, 4313 (1998).
\bibitem{bistability} M. Gurioli, L. Cavigli, G. Khitrova, and H. Gibbs,
 Phys. Stat. Sol. (a) {\bf 201}, 661 (2004).
\bibitem{wl} Q.Q. Wang, A. Muller, P. Bianucci, E. Rossi, Q.K. Xue,
 T. Takagahara, C. Piermarocchi, A. H. MacDonald, and C.K. Shih,
 http://arxiv.org/cond-mat/0404465.
\end{thebibliography}
\end{document}